\documentclass[12pt]{iopart}

\usepackage{inputenc}
\usepackage{hyperref}
\usepackage[T1]{fontenc}
\expandafter\let\csname equation*\endcsname\relax
\expandafter\let\csname endequation*\endcsname\relax
\usepackage{amsmath}
\usepackage{amsthm}
\usepackage{amssymb}
\usepackage{amsfonts}
\usepackage[english]{babel}
\usepackage{latexsym}
\usepackage{ulem}
\usepackage{xcolor}
\usepackage{txfonts}
\usepackage{pxfonts}
\usepackage{dsfont}
\usepackage{mathrsfs}
\usepackage{enumitem}
\usepackage{indentfirst}

\newcommand {\secref}[1] {(\ref{#1})}
\newcommand{\F}{\mathscr{F}}

\newcommand{\M}{\mathcal{M}}
\newcommand{\D}{\mathcal{D}}
\newcommand{\K}{\mathsf{K}}

\newtheorem{mydef}{Definition}[subsection]
\newtheorem{Lemma}{Lemma}[subsection]

\begin{document}

\title{States of Low Energy in Homogeneous and Inhomogeneous, Expanding Spacetimes}
\author{Kolja Them$^{1,2}$, Marcos Brum$^{3,1}$}
\address{$^{1}$ II. Institut f\"ur theoretische Physik, Universit\"at Hamburg -- Luruper Chaussee 149, D-22761 Hamburg (HH), Germany}
\address{$^{2}$ Present Address: Institute of Applied Physics and Microstructure Research Center, University of Hamburg -- Jungiusstrasse 11, D-20355 Hamburg (HH), Germany}
\ead{\mailto{kthem@physnet.uni-hamburg.de}}
\address{$^{3}$ Present Address: Instituto de F\'{\i}sica, Universidade Federal do Rio de Janeiro, Caixa Postal 68528, Rio de Janeiro, RJ 21941-972, Brazil}
\eads{\mailto{mbrum@if.ufrj.br}, \mailto{marcos.brum@desy.de}}

\begin{abstract}
We construct states on the algebra of the Klein-Gordon field that minimize the energy density in homogeneous and in inhomogeneous spacetimes, both with compact Cauchy hypersurfaces. The energy density is measured by geodesic observers and smeared over a spacelike slab of spacetime, entirely containing a Cauchy hypersurface and extended in time. We further show that these states are Hadamard states. The present construction generalizes the construction of States of Low Energy in Robertson-Walker spacetimes presented by Olbermann \cite{Olbermann07}.
\end{abstract}

\submitto{\CQG}

\maketitle

\section{Introduction}

In the algebraic approach, the quantum field theory is formulated from the assignment of an algebra of observables, chosen here to be the algebra of fields --- the solutions of the Klein-Gordon equation (see section \secref{subsecfieldquant} for more details). This approach becomes more advantageous in curved spacetimes, because of the absence of a well-defined concept of vacuum state. The incorporation of interacting fields into this framework  can be done perturbatively \cite{BruFreKoe96,BruFre00,HoWa02}.

Although the determination of a state space is not a cornerstone of the theory, states are needed to compute expectation values. Their construction in explicit cases is not a trivial task. On the other hand, since the solutions of the equations of motion are, in general, distributions, the analysis of their singularity structure is of highest importance. In this work we will focus on states which are completely described by their two-point functions, the so-called {\it quasifree states} (see section \eqref{secstates}). Among such states, the physically sensible ones are the so-called {\it Hadamard states}, for which a rigorous definition was first given in \cite{KayWald91}. Afterwards, it was shown that an equivalent definition can be given in terms of the wavefront set of the corresponding two-point function \cite{Radzikowski96,RadzikowskiVerch96}. This characterization is reminiscent of the spectral condition in Minkowski spacetime. It has also been shown that the expectation value of the energy-momentum tensor on these states is coherently renormalizable \cite{Wald94,Moretti03}. For the case of non-quasifree states, the singularities of all $n$-point functions can be determined from the singularities of their two-point function and the canonical commutation relations \cite{Sanders10}.

Parker \cite{Parker69} constructed states on asymptotically flat Robertson-Walker spacetimes, aiming at minimizing the production of asymptotically free particles, the so called {\it adiabatic states}. Neither was the asymptotic freedom verified, nor was this construction free of mathematical problems. Later on, L\"uders and Roberts \cite{LuRo90} presented a mathematically sound definition of adiabatic states and proved that any two adiabatic states are locally quasiequivalent. Afterwards, it was shown that all quasifree Hadamard states form a local quasiequivalence class\footnote{In the particular case of a globally hyperbolic spacetime with a compact Cauchy hypersurface, the set of all Hadamard quasifree states forms a unitary equivalence class \cite{Wald94}.} \cite{Verch94,BarGinouxPfaffle07}. More recently, Junker and Schrohe \cite{JunSchrohe02} extended the definition of adiabatic states for a certain class of globally hyperbolic spacetimes and showed that, under suitable conditions, the adiabatic states are Hadamard states.

In spite of the fact that the expectation value of the energy-momentum tensor of a scalar field on a Hadamard state is coherently renormalizable, it possesses no lower bound \cite{EpsGlaJaffe65}. This would give rise to violations of the second law of thermodynamics, as well as pathological spacetimes (allowing violations of causality). On the other hand, if instead of calculating the renormalized energy density at a particular point of spacetime, one smears it with the square of a smooth test function of compact support along the worldline of a causal observer, one finds that the resulting quantity cannot be arbitrarily negative. It was more recently shown that this result can also be obtained by smearing over a spacelike submanifold of spacetime. These results are known as Quantum Energy Inequalities (QEIs) \cite{Fewster00,FewsterSmith08}.

Inspired by these results, Olbermann \cite{Olbermann07} constructed states for the field algebra on RW spacetimes which are invariant under their symmetries and minimize the expectation value of the smeared renormalized energy density. Besides, using the adiabatic ansatz as initial condition for the scalar field and analyzing the singularity structure of those states, he was able to show that the States of Low Energy (SLE) are Hadamard States.

In this work, we generalize this construction in two different ways. First, we treat spacetimes whose Cauchy hypersurfaces are homogeneous and compact without boundary and show how to define symmetric states on such spacetimes. The set of spacetimes with homogeneous Cauchy hypersurfaces encompasses the well known Bianchi spacetimes \cite{Stephani03}. Hadamard states were recently constructed on these spacetimes by the authors of \cite{AvetisyanVerch12}. We will briefly comment on examples of spacetimes whose Cauchy hypersurfaces are compact and homogeneous. Second, we consider expanding spacetimes whose Cauchy hypersurfaces have no spatial symmetries but are compact and have no boundary. On both cases, we smear the energy density with respect to geodesic observers and prove that one can choose states which both minimize the energy density and satisfy the Hadamard condition. Besides, it will become clear that the latter case generalizes the former.

On section \secref{fieldquant} we present the scalar field quantization according to the algebraic approach. On section \secref{SLE-est} we introduce our generalization of the concept of States of Low Energy and, on section \secref{SLE-Hadamard}, we prove that the new states are also Hadamard states.

\section{Scalar Field quantization on Globally Hyperbolic Spacetimes}\label{fieldquant}

\subsection{Quantized scalar field}\label{subsecfieldquant}$ $

Globally hyperbolic spacetimes $\M$ are smooth, orientable, time orientable and paracompact manifolds, also possessing smooth Cauchy hypersurfaces \cite{BernalSanchez03,Wald84}. They have the topological structure $\M = \mathbb{R}\times\Sigma$ and the property that, for any point $p\in \M$, every inextendible causal curve through $p$ intersects $\Sigma$ exactly once. Hence, the determination of the solution of the equations of motion on $\Sigma$ fixes uniquely the field configuration at any point of spacetime \cite{Wald94}.

It is well known \cite{BarGinouxPfaffle07} that the wave equation of a massive scalar field on such spacetimes admits unique retarded and advanced fundamental solutions, which are maps $\mathds{E}^{\pm}:C_{0}^{\infty}(\M)\rightarrow C^{\infty}(\M)$, such that, for $f\in C_{0}^{\infty}(\M)\eqqcolon \D(\M)$,
\begin{equation}
 \left(\Box +m^{2}\right)\mathds{E}^{\pm}f=\mathds{E}^{\pm}\left(\Box +m^{2}\right)f=f
 \label{KGfund}
\end{equation}
and
\[\textrm{supp}(\mathds{E}^{\pm}f)\subset J^{\pm}(\textrm{supp}f) \; .\]
Here, $J^{+(-)}(A)$, for any subset $A\subset\M$, indicates the causal future (past) of $A$, i.e., the set of all points on $\M$ which can be reached from $A$ along a future-(past-)directed causal curve \cite{Wald84}.

The functions $f\in \D(\M)$ are called test functions, and $P\coloneqq \Box +m^{2}$ will denote the differential operator. From the fundamental solutions, one defines the {\it advanced-minus-retarded-operator} $\mathds{E}\coloneqq \mathds{E}^{-}-\mathds{E}^{+}$ as a map $\mathds{E}:C_{0}^{\infty}(\M)\rightarrow C^{\infty}(\M)$. Using $\mathds{E}$, we define the antisymmetric form
\begin{equation}
 \sigma(f,f')=-\int\textrm{d}^{4}x\sqrt{|g|}\, f(x)(\mathds{E}f')(x) = -E(f,f') \; .
 \label{symplform}
\end{equation}
This antisymmetric form is degenerate, because if $f_{1}$ and $f_{2}$, both elements of $\D(\M)$, are related by $f_{1}=Pf_{2}$, then $\forall f \in \D(\M)$ we have
\[\sigma(f,f_{1})=0 \; .\]
Therefore the domain of the antisymmetric form must be replaced by the quotient space $\D(\M)/\textrm{Ran}P\eqqcolon \K(\M)$\footnote{$\textrm{Ran}P$ is the {\it range} of the operator $P$, that is, the elements $f\in\D(\M)$ such that $f=Ph$ for some $h\in\D(\M)$.}. The pair $\left(\textrm{Re}(\K(\M)),\sigma\right)$ forms a symplectic vector space.

We now construct the algebra of fields. To each $f\in \K(\M)$ we assign the abstract symbol $\Phi(f)$ and construct the universal tensor algebra:
\[\mathscr{A} \coloneqq \bigoplus_{n=0}^{\infty}\K(\M)^{\otimes n} \; ,\]
where $\K(\M)^{(0)} \equiv \mathds{1}$. Endowing this algebra with a complex conjugation as a $^{*}$-operation and taking its quotient with the closed two-sided $^{*}$-ideal $\mathscr{J}$ generated by:
\[\Phi(f)\Phi(f')-\Phi(f')\Phi(f)+i\sigma(f,f')\mathds{1} \; ;\]
\[\Phi(Pf)=0 \; ,\]
we obtain a unital $^{*}$-algebra $\F$, the $CCR$-algebra. The symbols $\Phi(f)$ are distributions:
\begin{equation}
\Phi(f)=\int\textrm{d}^{4}x\sqrt{|g|}\, \phi(x)f(x) \; .
\label{field-distr}
\end{equation}
Moreover, a topology can be assigned to this algebra \cite{ArakiYamagami82}, turning $\F$ into a topological, unital $^{*}$-algebra.

Dimock \cite{Dimock80} showed that the $CCR$-algebra can be equivalently constructed using the initial-value fields, by setting $\phi = \mathds{E}f$ and $\psi = \mathds{E}f'$, where $f$ and $f'$ are test functions: one defines the restriction operators $\rho_{0}: \phi \mapsto \phi_{\upharpoonright \Sigma} \eqqcolon \phi_{0}$ and $\rho_{1}: \phi \mapsto (\partial_{n}\phi)_{\upharpoonright \Sigma} \eqqcolon \phi_{1}$ (and similarly for $\psi$), where $\partial_{n}$ is the derivative in the direction of the vector $n$, normal to $\Sigma$. The new space of functions is given by
\begin{equation}
 L(\Sigma)=\left\{(\phi_{0},\phi_{1})\in C_{0}^{\infty}(\Sigma)\times C_{0}^{\infty}(\Sigma)\right\} \; ,
 \label{Cauchysymplspace}
\end{equation}
and the symplectic form, by
\begin{equation}
 \sigma(f,f')=-\int_{\Sigma}\textrm{d}^{3}x\sqrt{|g_{\upharpoonright \Sigma}|}\, \left(\overline{\phi}_{0}(x)\psi_{1}(x)-\psi_{0}(x)\overline{\phi}_{1}(x)\right) \; .
 \label{Cauchysymplform}
\end{equation}
The symplectic form defined above does not dependend on the Cauchy hypersurface on which it is calculated and it is preserved by the isomorphic mapping $\beta:K\rightarrow L\textrm{\ ,\ }\phi\mapsto (\phi_{1},\phi_{2})$.

On a general globally hyperbolic spacetime, one can always choose a coordinate system on which the metric takes the form \cite{BarGinouxPfaffle07}
\begin{equation}
 ds^{2}=\Gamma dt^{2}-h_{t} \; ,
 \label{metric-GH}
\end{equation}
where $\Gamma$ is a positive smooth function and $h_{t}$ is a Riemannian metric on $\Sigma$ depending smoothly on $t\in\mathbb{R}$. But the Klein-Gordon equation arising from such a metric is not, in general, separable. We will make here the assumptions that $\Gamma\equiv 1$ and that the metric on the spatial hypersurfaces can be written in the following form:
\begin{equation}
 h_{t}=c(t)^{2}h_{ij}(\uline{x})d\uline{x}^{i}d\uline{x}^{j}\; .
 \label{KGansatz} 
\end{equation}
We will call such spacetimes {\it expanding spacetimes}. Here, $c(t)$ is a smooth positive function of time, the so called {\it scale factor}, and $h_{ij}(\uline{x})$ is the metric on the Riemannian hypersurfaces ($\uline{x}$ denotes the spatial coordinates of a point on the manifold). The metric assumes the usual form
\begin{equation}
 ds^{2}=dt^{2}-c(t)^{2}h_{ij}(\uline{x})d\uline{x}^{i}d\uline{x}^{j}\; ,
 \label{metric-timedecomp-GH}
\end{equation}
and Klein-Gordon equation for a scalar field $\phi(t,\uline{x})$ assumes, then, the form
\begin{equation}
 \left(\partial_{t}^{2}+3\frac{\dot{c}(t)}{c(t)}\partial_{t}-\frac{\Delta_{h}}{c(t)^{2}}+m^{2}\right)\phi(t,\uline{x})=0 \; .
 \label{KG}
\end{equation}

If the Riemannian hypersurfaces are compact, which will be the case in the problems treated below\footnote{The problem of mode decomposition for spacetimes with noncompact homogeneous Riemannian hypersurfaces was treated in \cite{Avetisyan12}. We are indebted to the author of that paper for stressing the validity of our treatment.}, the Laplace operator $-\Delta_{h}$ becomes an essentially self-adjoint operator on the Hilbert space $L^{2}(\Sigma,\sqrt{|h|})$. We will also denote its unique self-adjoint extension by $-\Delta_{h}$. This extension possesses a complete set of orthonormal eigenfunctions $\psi_{j}$ (on the following, we will, for short, refer to $\psi_{j}$ as eigenfunctions of the laplacian) and the index $j$ runs over a countable set $I$. The corresponding eigenvalues $\lambda_{j}$ form a discrete spectrum and they form a nondecreasing sequence, i.e., for $j_{1}>j_{2}$, $\lambda_{j_{1}}>\lambda_{j_{2}}$ \cite{Jost11}. Also, $\forall j\neq 0\ ,\ \lambda_{j}>0$. The solution to equation \eqref{KG} can be written as
\begin{equation}
 \phi(t,\uline{x})=T_{j}(t)\overline{\psi}_{j}(\uline{x}) \; ,
 \label{KGfield}
\end{equation}
and $T_{j}$ must satisfy
\begin{equation}
 \left(\partial_{t}^{2}+3\frac{\dot{c}(t)}{c(t)}\partial_{t}+\omega_{j}^{2}(t)\right)T_{j}(t)=0 \; ,
 \label{timeKG}
\end{equation}
where
\begin{equation}
\omega_{j}^{2}(t)\coloneqq\frac{\lambda_{j}}{c(t)^{2}}+m^{2} \; .
\label{frequency}
\end{equation}

The two linearly independent real valued solutions of equation \eqref{timeKG} can be combined in a complex valued solution which satisfies the normalization condition
\begin{equation}
\dot{T}_{j}(t)\overline{T}_{j}(t)-T_{j}(t)\dot{\overline{T}}_{j}(t)=\frac{i}{c(t)^{3}} \; .
\label{timeSymplprod}
\end{equation}
Since the left-hand side is the Wronskian $W[T_{j},\overline{T}_{j}]$, $T_{j}(t)$ and $\overline{T}_{j}(t)$ are linearly independent. From this linear independence, if $S_{j}(t)$ and $\overline{S}_{j}(t)$ are also linearly independent solutions of \eqref{timeKG}, we can write
\begin{equation}
 T_{j}(t)=\alpha_{j}S_{j}(t)+\beta_{j}\overline{S}_{j}(t)\; .
 \label{Bogtransf}
\end{equation}
Since $S_{j}(t)$ must also satisfy \eqref{timeSymplprod}, the parameters $\alpha_{j}$ and $\beta_{j}$ are then subject to
\begin{equation}
 |\alpha_{j}|^{2}-|\beta_{j}|^{2}=1\; .
 \label{Bogcoef}
\end{equation}
The solutions to equation \eqref{timeKG} have only two free parameters. On the other hand, taking into account the absolute values and phases of $\alpha_{j}$ and $\beta_{j}$, subject to \eqref{Bogcoef}, we would have three free parameters. One of these is then a free parameter. Throughout this paper, we choose $\beta_{j}$ to be a real parameter.

\subsection{States and the Hadamard condition}\label{secstates}$ $

States $\omega$ are functionals over the $CCR$-algebra $\F$ with the following properties\footnote{We remind the reader that the algebra $\F$ is a unital topological $^{*}$-algebra. Otherwise, the definition of state would not be correct.}: 
\begin{description}
 \item [Linearity] $\omega(\alpha A+\beta B)=\alpha\omega(A)+\beta\omega(B)$, $\alpha$, $\beta\in \mathbb{C}$, $A$, $B\in \F$;
 \item [Positive-semidefiniteness] $\omega(A^{*}A)\geq 0$;
 \item [Normalization] $\omega(\mathds{1})=1$.
\end{description}
The $n$-{\it point functions} of $\omega$ are defined as
\[w_{\omega}^{(n)}(f_{1}\otimes \ldots \otimes f_{n}) \coloneqq \omega(\Phi(f_{1})\ldots\Phi(f_{n})) \; .\]
The higher-point functions also possess the linearity property, i.e., $w_{\omega}^{(n)}:\F^{\otimes n} \rightarrow \mathbb{C}$ is linear in each of its arguments.

In the present work we will focus on states which are completely described by their two-point functions, the so-called {\it quasifree states}. All odd-point functions vanish identically and the higher even-point functions can be written as
\[w^{(2n)}_{\omega}(f_{1} \otimes \ldots \otimes f_{2n}) = \sum_{p}\prod_{k=1}^{n} w^{(2)}_{\omega}(f_{p(k)},f_{p(k+n)})\; .\]
Here, $w^{(2n)}_{\omega}$ is the $2n$-point function associated to the state $\omega$, $w^{(2)}_{\omega}(f_{p(k)},f_{p(k+n)})\equiv\omega(f_{p(k)},f_{p(k+n)})$ and the sum runs over all permutations of $\{1,\ldots ,n\}$ which satisfy $p(1)< \ldots <p(n)$ and $p(k)<p(k+n)$. We call a state {\it pure} if it is not a convex combination of two distinct states, i.e.,
\[\nexists \, \omega_{1},\omega_{2}\, \textrm{distinct states over }\F,\, \, \textrm{and }\lambda \in (0,1)\, \, |\, \omega=\lambda\omega_{1}+(1-\lambda)\omega_{2} \; .\]

The existence of representations of the fields as operators on a certain Hilbert space is achieved by means of the GNS construction \cite{BraRob-I}: given a $^{*}$-algebra $\F$ and a state $\omega$ over this algebra\footnote{One can start with a positive semi-definite functional and take the quotient of the algebra by the left ideal given by
\[\mathcal{N}_{\omega}=\left\{F\, ;F \in \F\; ,\, \omega(F^{*}F)=0\right\} \; .\]}, there exists a representation $\left(\mathscr{H}_{\omega},\pi_{\omega}\right)$ consisting of a Hilbert space $\mathscr{H}_{\omega}$ and a representation $\pi_{\omega}$ of the algebra $\F$ as operators on this Hilbert space. Also, $\exists \Omega\in\mathscr{H}_{\omega}$ such that, $\forall F\in \F$, $\omega(F)=\langle\Omega|\pi_{\omega}(F)|\Omega\rangle$, and $\Omega$ is a cyclic vector in $\mathscr{H}_{\omega}$. The triple $\left(\mathscr{H}_{\omega},\pi_{\omega},\Omega\right)$ is unique up to unitary equivalence and the representation is irreducible if and only if the state $\omega$ is pure.

The physically meaningful states are the Hadamard states, which are characterized by the singularity structure of their two-point function. This characterization is reminiscent of the spectral condition on Minkowski spacetime. The formal definition of Hadamard states (see below) is given in terms of the wavefront set $WF$ of the two-point distributions corresponding to the states \cite{Radzikowski96,RadzikowskiVerch96,Hormander-I}. It is also known that the expectation value of the energy-momentum tensor on these states has a sensible renormalization \cite{Wald94,Moretti03}. The advantage of this approach is that the $WF$ is a geometrical object and therefore the location of singularities can be coherently analyzed, even in curved spacetimes. Besides, it allows the incorporation of interacting field theories at the perturbative level in the algebraic approach \cite{BruFreKoe96,BruFre00,HoWa02}. We will present this definition now.

Let $v$ be a distribution of compact support. If $\forall N \in \mathbb{N}_{0} \; ,\, \exists C_{N} \in \mathbb{R}_{+}$ such that
\begin{equation}
 \lvert \hat{v}(k) \rvert \leqslant C_{N}\left(1+\lvert k \rvert\right)^{-N} \; ,\, k \in \mathbb{R}^{n} \; ,
 \label{regsupp}
\end{equation}
then $v$ is in $C_{0}^{\infty}(\mathbb{R}^{n})$. Accordingly, the singular support ({\it singsupp}) of $v$ is defined as the set of points having no neighborhood where $v$ is in $C^{\infty}$. Moreover, we define the cone $\Sigma(v)$ as the set of points $k \in \mathbb{R}^{n}\diagdown\{0\}$ having no conic neighborhood $V$ such that \eqref{regsupp} is valid when $k \in V$.

For a general distribution $u \in \D '(X)$, where $X$ is an open set in $\mathbb{R}^{n}$ and $\phi \in C_{0}^{\infty}(X)$, $\phi(x)\neq 0$, we define
\[\Sigma_{x}(u) \coloneqq \bigcap_{\phi}\Sigma(\phi u) \; .\]

\begin{mydef}\label{smooth-wf}
 If $u \in \D '(X)$ then the {\it Wavefront set} of $u$ is the closed subset of $X \times (\mathbb{R}^{n}\diagdown\{0\})$ defined by
\[WF(u)=\{(x,k) \in X \times (\mathbb{R}^{n}\diagdown\{0\}) \arrowvert \, x \in \textrm{\it singsupp }u \, ,\, k \in \Sigma_{x}(u)\} \; .\]
\end{mydef}
In \cite{Hormander-I} it was proved that the wavefront set of a distribution defined on $\mathbb{R}^{n}$ transforms under change of coordinates as an element of the cotangent bundle ${\mathcal T}^{*}\mathbb{R}^{n}$. In \cite{FewsterSmith08} the authors remarked that, if $u$ is a distribution on the $m$-dimensional manifold $\M$, $(x,k)\in WF(u)\subset{\mathcal T}^{*}\M \diagdown \{0\}$ if and only if there exists a chart neighborhood $(\kappa,{\mathcal U})$ of $x$ such that the corresponding coordinate expression of $(x,k)$ belongs to $WF(u\circ\kappa^{-1})\subset \mathbb{R}^{m}\times\mathbb{R}^{m} \diagdown \{0\}$. Besides, the wavefront set of $u$ is independent of the particular chart chosen.

Finally, quasifree Hadamard states are defined by the following
\begin{mydef}\label{Hadamard-wf}
 A quasifree state $\omega$ is said to be a {\it Hadamard state} if its two-point distribution $\omega_{2}$ has the following Wavefront set:
\begin{equation}
 WF(\omega_{2})=\left\{\left(x_{1},k_{1};x_{2},-k_{2}\right) | \left(x_{1},k_{1};x_{2},k_{2}\right)\in {\mathcal T}^{*}\left(\M\times\M\right) \diagdown \{0\} ; (x_{1},k_{1})\sim (x_{2},k_{2}) ; k_{1}\in \overline{V}_{+}\right\}
 \label{Wfcond}
\end{equation}
where $(x_{1},k_{1})\sim (x_{2},k_{2})$ means that there exists a null geodesic connecting $x_{1}$ and $x_{2}$, $k_{1}$ is the cotangent vector to this geodesic at $x_{1}$ and $k_{2}$, its parallel transport, along this geodesic, at $x_{2}$. $\overline{V}_{+}$ is the closed forward light cone of $\mathcal{T}^{*}_{x_{1}}\M$.
\end{mydef}

To facilitate the writing, we will call this set $C^{+}$ and say that a quasifree state is Hadamard if its two-point function has this wavefront set:
\begin{equation}
WF(\omega_{2})=C^{+}\; .
\label{HadWF-C+}
\end{equation}
As stated in the introduction, the Hadamard condition can be similarly formulated for non-quasifree states \cite{Sanders10}.

One useful property of $WF$, which will be used later, is that for two distributions $\phi$ and $\psi$,
\begin{equation}
 WF(\phi +\psi)\subseteq WF(\phi) \cup WF(\psi) \; .
 \label{WFsum} 
\end{equation}
If the $WF$ of one of the distributions is empty, i.e., if one of them is smooth, then this inclusion becomes an equality.

Finally, we remark that the Hadamard condition only imposes a restriction on the singular structure of the two-point function of a state, the smooth part of the two-point function remaining completely undetermined. Besides, since states are linear functionals on the algebra of fields, it is immediate to see that the antisymmetric part of the two-point function of a quasifree state coincides with the causal propagator associated to the Klein-Gordon operator (up to a multiplicative factor of {\it i}).

\subsection{Quasifree states in Homogeneous spacetimes}$ $

The spatial hypersurfaces are Riemannian submanifolds, and we will now present the definition of homogeneity on them.

\begin{mydef}
 Let $G$ be a group of isometries from the Riemannian manifold $\Sigma$ to itself, i.e., $g\in G$ is a diffeomorphism from $\Sigma$ to itself and $\forall g\in G$, $g^{*}h=h$, where $h$ is the metric on $\Sigma$. If for every pair of points $p,q\in\Sigma\textrm{, }\exists g'\in G$ such that $g'p=q$, then the group $G$ is said to act transitively on $\Sigma$. A Riemannian manifold with a transitive group of isometries is called {\it homogeneous} \cite{Jost11}.
\end{mydef}

The action of the group $G$ as a group of isometries at a point $(t,x)\in\M$ is $g(t,x)=(t,gx)$, where $t\in\mathbb{R}$ and $x\in\Sigma$. The homogeneous spaces can be classified according to their Lie-group structure \cite{Osinovsky73} and are designated as Bianchi I-IX spaces. From this classification, one can construct globally hyperbolic spacetimes whose Cauchy hypersurfaces are isometric to one of those homogeneous spaces. Such spacetimes are called {\it Bianchi spacetimes}.

We are interested in spacetimes with compact Riemannian hypersurfaces without boundary because, in this case, each eigenvalue of the laplacian has finite multiplicity \cite{Berard86}, thus simplifying both the mode decomposition presented earlier and the construction of symmetric states.  The Bianchi spaces I-VIII are topologically equivalent to $\mathbb{R}^{3}$, therefore noncompact. The symmetry structure of the Bianchi IX space is given by the SU(2) group, which is already compact. Among the noncompact ones, the simplest is Bianchi I, which has a commutative group structure. By taking the quotient between this group and the group $\mathbb{Z}$ of integer numbers, the resulting space is the 3-torus. This is a compact space without boundary. We remark that there exists more than one method of compactification (see \cite{Tanimoto04} and references therein).

$G$ has a unitary representation $U\oplus U$ on $L^2(\Sigma)\oplus L^2(\Sigma)$, given by $U(g)f=f\circ g^{-1}$, such that
\begin{equation}
 \alpha_{g}\left(\phi(f)\right) = \phi(f\circ g^{-1}) \; .
 \label{automorph}
\end{equation}
A quasifree state $\omega$ is said to be symmetric if $\forall g\in G$, $\omega \circ \alpha_{g} =\omega$. The quasifree symmetric state will be denoted by $\omega_{G}$.

The Riemannian metric on $\Sigma$ induces a scalar product on $L^2(\Sigma)\oplus L^2(\Sigma)$. Working with the initial-value fields $F=\left(F_{0},F_{1}\right)=\left(\rho_{0}\mathds{E}f,\rho_{1}\mathds{E}f\right)$ and $F'=\left(F'_{0},F'_{1}\right)=\left(\rho_{0}\mathds{E}f',\rho_{1}\mathds{E}f'\right)$, we have
\begin{equation}
 \left(F,F'\right)=\int\textrm{d}^{3}x\sqrt{|h|}\, \left(\overline{F}_{0}F'_{0}+\overline{F}_{1}F'_{1}\right) \; .
\end{equation}

Schwarz's nuclear theorem states that to the two-point function $S$ in the space of initial-value fields is associated an element of the dual to $L^2(\Sigma)$:
\[L^2(\Sigma)\ni F' \, \mapsto \, S(\cdot ,F')\in \left(L^2(\Sigma)\right)^{*} \; .\]
Now, from Riesz's representation theorem, to the element $S(\cdot ,F')\in \left(L^2(\Sigma)\right)^{*}$ there exists associated an element $\hat{S}(F')\in L^2(\Sigma)$

Therefore, $\forall F\in L^2(\Sigma)$,
\begin{equation}
 S(F,F')=\left(F,\hat{S}(F')\right) \; .
 \label{twoptriesz}
\end{equation}

Using the eigenfunctions of the laplacian, a generalized Fourier transform can be defined:
\begin{equation}
 \widetilde{F}_{j}:=\left(\psi_{j},F\right)_{L^2}=\int \textrm{d}^{3}x\sqrt{|h|}\, \overline{\psi}_{j}(\uline{x})F(x) \; .
 \label{Fouriertransf}
\end{equation}
The fact that the $\psi_{j}$ form a complete basis of orthonormal eigenfunctions of the laplacian operator allows us to write
\begin{equation}
 S\left(F,F'\right)=\sum_{j}\langle \widetilde{\overline{F}}_{j},\widetilde{\hat{S} (F')}_{j} \rangle \; ,
\end{equation}
where
\begin{equation}
 \langle \widetilde{\overline{F}}_{j},\widetilde{\hat{S} (F')}_{j} \rangle =\sum_{l=0}^{1}\widetilde{\overline{F}_{l}}_{j}\left(\widetilde{\hat{S} (F')_{l}}\right)_{j} \; .
\end{equation}
This last sum is over $l$ ranging from $0$ to $1$ because $F$ and $F'$ are representing the initial-value fields $\left(F_{0},F_{1}\right)$ and $\left(F'_{0},F'_{1}\right)$.

The proof that this two-point function gives rise to a quasifree homogeneous state follows from \cite{LuRo90}. The only difference from the proof now is that, in RW spacetimes, the commutant of each symmetry group consists of diagonalizable operators (see Appendix A of that reference), i.e., operators $T$ such that
\[\left(\widetilde{Tf}\right)_{j}=t_{j}\widetilde{f}_{j} \; .\]
This will not be generally true in our case. There, this fact led to the conclusion that the operation of $\hat{S}$ on a test function, evaluated in Fourier space, simply amounted to a multiplication by a function of the mode, i.e., $\widetilde{\hat{S} (F')}_{j}=\widetilde{\hat{S}}_{j}\widetilde{F'}_{j}$, but this is not true here. Nevertheless the same analysis made there for the operator $\widetilde{\hat{S}}_{j}$ can be made here for $\widetilde{\hat{S} (F')}_{j}$. In our case, this results in the construction of a quasifree homogeneous state, while there the state was also isotropic.

The two-point functions of the homogeneous states are given by
\begin{equation}
 w^{(2)}_{\omega_{G}}(x,x')=\sum_{j}\overline{T}_{j}(t)T_{j}(t')\psi_{j}(\uline{x})\overline{\psi}_{j}(\uline{x'})
 \label{2ptfcnHom}
\end{equation}
and $T_{j}$ has initial conditions at time $t_{0}$ given by
\[T_{j}(t_{0})=q_{j} \; \; , \; \; \dot{T}_{j}(t_{0})=c^{-3}(t_{0})p_{j} \; ,\]
where $q_{j}$ and $p_{j}$ are polynomially bounded functions. Elliptic regularity guarantees the boundedness of $\psi_{j}(\uline{x})\overline{\psi}_{j}(\uline{x'})$.

\subsection{Quasifree states in Expanding Spacetimes}$ $

In this subsection, we are going to construct quasifree states in spacetimes without spatial symmetries. Therefore, the discussion about symmetric states becomes meaningless here. However, we will now show that the GNS construction presented in section \secref{secstates} provides us states whose two-point functions can be written as in \eqref{2ptfcnHom}.

Given a state $\omega$ and the corresponding cyclic vector $\Omega\in\mathscr{H}_{\omega}$, we will expand the representation of the field in terms of the operator $a$ and its adjoint, $a^{\dagger}$, such that \[a|\Omega\rangle=0\; .\]

On those spacetimes which we denoted {\it expanding spacetimes} (see section \secref{subsecfieldquant}), the KG operator separates as a laplacian operator on the Cauchy hypersurfaces and an ordinary differential operator (see \eqref{timeKG}). Hence, the field can be expressed as (we use the symbol $\phi$ to denote both the field and its representation on the Hilbert space generated by the GNS construction)
\begin{equation}
 \phi(t,\uline{x})=\frac{1}{\sqrt{2}}\left[a_{j}\overline{T}_{j}(t)\psi_{j}(\uline{x})+a^{\dagger}_{j}T_{j}(t)\overline {\psi}_{j}(\uline{x})\right]\; .
 \label{fieldrepr}
\end{equation}
The operator $a$ and its adjoint also follow the mode decomposition. These operators satisfy the usual commutation relations: \[\Big[a_{j},a_{j'}\Big]=\Big[a^{\dagger}_{j},a^{\dagger}_{j'}\Big]=0\textrm{\quad} \Big[a_{j},a^{\dagger}_{j'}\Big]=\delta_{jj'}\; ,\]
where $\delta_{jj'}$ is the Kronecker delta.

Evaluated on the state $|\Omega\rangle$, the two-point function of this field operator is ($f$ and $f'$ are test functions of compact support)
\begin{align}
 w^{(2)}_{\omega}\left(\phi(f)\phi(f')\right) &=\int\textrm{d}^{4}x\sqrt{|g(x)|}\textrm{d}^{4}x'\sqrt{|g(x')|}\, f(t,\uline{x})f'(s,\uline{x'})\sum_{j}\overline{T}_{j}(t)T_{j}(s)\psi_{j}(\uline{x})\overline{\psi}_{j}(\uline{x'})\; . \nonumber \\
 &\eqqcolon\int\textrm{d}^{4}x\sqrt{|g(x)|}\textrm{d}^{4}x'\sqrt{|g(x')|}\, f(t,\uline{x})f'(s,\uline{x'})w^{(2)}_{\omega}(\uline{x},\uline{x'}) \; .
 \label{GNS2ptfcn}
\end{align}
Here we note that this two-point function has the same form as \eqref{2ptfcnHom}, but it was formulated for general expanding spacetimes, therefore it has a wider range of aplicability than the former one.

Regarding the convergence of the sum in equation \eqref{GNS2ptfcn}, we remark that we will write $T_{j}$ in the form \eqref{Bogtransf} with $\alpha$ and $\beta$ subject to \eqref{Bogcoef} and $\beta$ chosen to be real. For the solution $S_{j}(t)$ we will choose initial conditions given by the $N$-fold iteration of the adiabatic ansatz --- see section \secref{subsecfulfillhadamard}. There we will show that, for a large $N$, the state whose two-point function is given by \eqref{GNS2ptfcn} is a Hadamard state.

\section{States of Low Energy in Expanding Spacetimes}\label{SLE-est}

We will now construct the States of Low Energy in expanding spacetimes without spatial symmetries but with compact Cauchy hypersurface without boundary. We will show that the construction in homogeneous spacetimes is a particular case of the one presented in this section. We will point out the differences between our construction and the original one, given in \cite{Olbermann07}.

The renormalized energy density will be obtained by means of the point-splitting method. In the absence of spatial symmetries, this quantity must be dependent on position, therefore we will need to smear it over a spatially extended spacelike submanifold. Since the Cauchy hypersurfaces are compact, we can perform the smearing with test functions which do not depend on the spatial position. In the homogeneous case, the renormalized energy density will be integrated over a Cauchy hypersurface and the result will be divided by the volume of this region, as in the particular case of RW spacetime with positive spatial curvature \cite{Olbermann07}. We want to stress here that in the RW spacetimes with negative or null spatial curvature there was no need to perform the smearing in space, whereas it is necessary in the inhomogeneous case.

We will choose as observers a congruence of geodesic curves which are everywhere orthogonal to the Cauchy hypersurfaces. This means that for every such observer, its four velocity $\dot{\gamma}$ is orthogonal to every vector $X\in\mathcal{T}_{p}\Sigma$, for every point $p$ in $\Sigma$. In the coordinate system we have chosen, with metric of the form \eqref{metric-timedecomp-GH}, the ortogonality condition becomes
\begin{equation}
 g(\dot{\gamma},X)=-c^{2}(t)h_{kl}(\uline{x})\dot{\gamma}^{k}(t,\uline{x})X^{l}(t,\uline{x})=0\, \therefore \dot{\gamma}^{k}\equiv 0 \; .
\end{equation}
We also require that the four velocity is normalized. Thus,
\begin{equation}
 g(\dot{\gamma},\dot{\gamma})=(\dot{\gamma}^{0})^{2}=1\, \therefore \dot{\gamma}^{0}=1 \; .
\end{equation}

The energy density measured by the chosen observers is evaluated from the energy-momentum tensor $T_{ab}(x)$ as
\begin{equation}
 \rho(x)=T_{\mu\nu}(x)\dot{\gamma}^{\mu}\dot{\gamma}^{\nu}=T_{00}(x)\; .
\end{equation}

The expectation value of the regularized energy density on a state $\omega$ is
\begin{equation}
 \langle \hat{T}^{reg}\rangle_{\omega}(x,x')=\left[\frac{1}{2}\nabla_{0}|_{x}\nabla_{0}|_{x'}+\frac{1}{2}\nabla^{c}|_{x}\nabla_{c}|_{x'}+\frac{1}{2}m^{2}\right]w^{(2)}_{\omega}(x,x')\; .
 \label{Treg_u}
\end{equation}

The renormalized energy density is encountered by subtracting from this expression the expectation value of the energy density on a reference Hadamard state $\omega_{0}$,
\begin{equation}
 \langle \hat{T}^{ren}\rangle_{\omega}=\langle \hat{T}^{reg}\rangle_{\omega}-\langle \hat{T}^{reg}\rangle_{\omega_{0}} \; ,
\label{pointsplit-general}
\end{equation}
and then taking the coincidence limit $x'\rightarrow x$. The renormalization counterterms amount to purely geometrical terms \cite{Wald94,Moretti03,HackPhD}. Since these are independent of the state $\omega$ they turn out to be irrelevant for the determination of the SLE. Therefore, they will not be written in the following.

By performing the Bogolubov transformation \eqref{Bogtransf}, $w^{(2)}_{\omega}(x,x')$ becomes
\begin{align}
 w^{(2)}_{\omega}(x,x')=\sum_{j} &\left[(1+\beta_{j}^{2})\overline{S}_{j}(t)S_{j}(s)+\beta_{j}^{2}S_{j}(t)\overline{S}_{j}(s) \right. \nonumber \\
 &\left. +2\beta_{j}\sqrt{1+\beta_{j}^{2}}\textrm{Re}\left(e^{i\theta_{j}}S_{j}(t)S_{j}(s)\right)\right]\psi_{j}(\uline{x})\overline{\psi}_{j}(\uline{x'}) \nonumber \\
 &\eqqcolon \sum_{j}w^{(2)}_{\omega_{j}}(x,x') \; .
 \label{Bogstate}
\end{align}
where $\alpha_{j}=e^{i\theta_{j}}\sqrt{1+\beta_{j}^{2}}$ ($\beta_{j}$ was chosen to be real --- see remarks after equation \eqref{Bogcoef}). The last equality in \eqref{Bogstate} shows that the minimization amounts to finding the Bogolubov parameters $\beta_{j}$ and $\theta_{j}$ which minimize the contribution of each mode to the energy density. Since the last term in equation \eqref{pointsplit-general} is independent of the state $\omega$, and therefore independent of the Bogolubov parameters, this term becomes irrelevant for the present purposes. Regarding the convergence of the sum in equation \eqref{Bogstate}, see remarks after equation \eqref{GNS2ptfcn}.

The definition of $w^{(2)}_{\omega_{j}}$ allows us to make a mode decomposition of the expectation value of the regularized energy density on the state $\omega$. We thus define
\begin{equation}
 \langle \hat{T}^{reg}\rangle_{\omega_{j}}(x,x')\coloneqq\left[\frac{1}{2}\nabla_{0}|_{x}\nabla_{0}|_{x'}+\frac{1}{2}\nabla^{c}|_{x}\nabla_{c}|_{x'}+\frac{1}{2}m^{2}\right]w^{(2)}_{\omega_{j}}(x,x') \; .
 \label{Treg-mode}
\end{equation}
Inserting \eqref{Bogstate} into \eqref{Treg-mode} and taking the coincidence limit $x'\rightarrow x$, we find
\begin{align}
 &\langle \hat{T}\rangle_{\omega_{j}}(t,\uline{x}) \coloneqq \lim_{x'\rightarrow x}\langle \hat{T}^{reg}\rangle_{\omega_{j}}(x,x')= \nonumber \\
 &\frac{1}{2}(1+2\beta_{j}^{2})\left\{\lvert\dot{S}_{j}(t)\rvert^{2}\lvert \psi_{j}(\uline{x})\rvert^{2}+\lvert S_{j}(t)\rvert^{2}\left(c(t)^{-2}h^{kl}(\uline{x})\nabla_{k}\psi_{j}(\uline{x})\nabla_{l}\overline{\psi}_{j}(\uline{x})+m^{2}\lvert \psi_{j}(\uline{x})\rvert^{2}\right)\right\} \nonumber \\
 &+\frac{1}{2}2\beta_{j}\sqrt{1+\beta_{j}^{2}}\textrm{Re}\left\{e^{i\theta_{j}}\left[\left(\dot{S}_{j}(t)\right)^{2}\lvert \psi_{j}(\uline{x})\rvert^{2} \right. \right. \nonumber \\
 &\left. \left. +\left(S_{j}(t)\right)^{2}\left(c(t)^{-2}h^{kl}(\uline{x})\nabla_{k}\psi_{j}(\uline{x})\nabla_{l}\overline{\psi}_{j}(\uline{x})+m^{2}\lvert \psi_{j}(\uline{x})\rvert^{2}\right)\right]\right\}\; .
 \label{EnergyGNS}
\end{align}

It is clear from equation \eqref{EnergyGNS} that, if the energy density is not smeared also in space, the parameters $\beta_{j}$ and $\theta_{j}$ will not be constants. The smeared energy density is now
\begin{equation}
 E_{j} \coloneqq \int_{\mathbb{R}}\textrm{d}t\, f^{2}(t)\int_{\Sigma}\textrm{d}^{3}x\sqrt{|h|}\, \langle \hat{T}\rangle_{\omega_{j}}(t,\uline{x}) \; .
\end{equation}
This should be interpreted as a heuristic formula, since this is just the coincidence limit of the expectation value of the regularized energy density, not the renormalized one. However, as stated above, this is the term which must be analyzed in order to construct the SLE.

Since the spatial hypersurfaces are compact without boundary, we calculate
\begin{align}
 \int_{\Sigma}\textrm{d}^{3}x\sqrt{|h|}\, h^{kl}(\uline{x})\nabla_{k}\psi_{j}(\uline{x})\nabla_{l}\overline{\psi}_{j}(\uline{x}) &=-\int_{\Sigma}\textrm{d}^{3}x\sqrt{|h|}\, \psi_{j}(\uline{x})\Delta_{h}\overline{\psi}_{j}(\uline{x}) \nonumber \\
 &=\lambda_{j}\int_{\Sigma}\textrm{d}^{3}x\sqrt{|h|}\, \lvert \psi_{j}(\uline{x})\rvert^{2}=\lambda_{j}\; .
 \label{laplcompsurfc}
\end{align}

Therefore,
\begin{align}
 E_{j} &=(1+2\beta_{j}^{2})\frac{1}{2}\int\textrm{d}t\, f^{2}(t)\left(|\dot{S}_{j}(t)|^{2}+\omega_{j}^{2}(t)|S_{j}(t)|^{2}\right) \nonumber \\
 &+2\beta_{j}\sqrt{1+\beta_{j}^{2}}\frac{1}{2}\textrm{Re}\left\{e^{i\theta_{j}}\int\textrm{d}t\, f^{2}(t)\left((\dot{S}_{j}(t))^{2}+\omega_{j}^{2}(t)S_{j}(t)^{2}\right)\right\} \nonumber \\
 &=(1+2\beta_{j}^{2})c_{1j}+2\beta_{j}\sqrt{1+\beta_{j}^{2}}\textrm{Re}(e^{i\theta_{j}}c_{2j})\; ,
 \label{EnergyGNSmin}
\end{align}
where
\begin{align}
&c_{1j}=\frac{1}{2}\int\textrm{d}t\, f^{2}(t)\left(|\dot{S}_{j}(t)|^{2}+\omega_{j}^{2}(t)|S_{j}(t)|^{2}\right) \label{c1j} \\
&c_{2j}=\frac{1}{2}\int\textrm{d}t\, f^{2}(t)\left((\dot{S}_{j}(t))^{2}+\omega_{j}^{2}(t)S_{j}(t)^{2}\right) \; . \label{c2j}
\end{align}
It is easy to see that, by choosing
\begin{equation}
 \beta_{j}=\sqrt{\frac{c_{1j}}{2\sqrt{c_{1j}^{2}-|c_{2j}|^{2}}}-\frac{1}{2}}\textrm{\quad and \quad}\alpha_{j}=e^{i\theta_{j}}\sqrt{\frac{c_{1j}}{2\sqrt{c_{1j}^{2}-|c_{2j}|^{2}}}+\frac{1}{2}}
 \label{Bogmin}
\end{equation}
and
\begin{equation}
\theta_{j}=-\textrm{Arg}c_{2j}+\pi\; ,
\label{thetamin}
\end{equation}
we minimize \eqref{EnergyGNSmin}. We will refer to these states of low energy as $\omega_{SLE}$, and their two-point functions will be referred to as $w^{(2)}_{\omega_{SLE}}$. The proof that these states are of the Hadamard form will be left for the next section.

We remark that the SLE are dependent on the test function used in the smearing. In spite of that, Degner \cite{Degner09} calculated the particle production process on such states in RW spacetimes and showed that the rate of production is not strongly dependent on the test function chosen. This dependence would only be dropped if the terms between parentheses in equations \eqref{c1j} and \eqref{c2j} could be taken out of the integrals. This would be the case if and only if $c(t)=$ constant, and in such a case we would have $c_{2j}\equiv 0$ and the SLE would reduce to the static vacuum.

The SLE constructed here are different from the ones constructed by Olbermann because here we minimize the energy density over a spacelike slab of spacetime (containing entirely a Cauchy hypersurface and extended in time), while there the integration over a spatially extended region was not in general necessary. Besides, the treatment given here does not depend on the occurrence of spatial symmetries. We also note that if we had chosen an arbitrary causal observer, the energy density would contain terms of the form $\dot{\gamma}^{0}\dot{\gamma}^{l}\nabla_{0}|_{x}\nabla_{l}|_{x'}w^{(2)}_{\omega}(x,x')$, which could spoil the positivity of \eqref{c1j}, thus compromising the minimization of the energy density. Such a problem would also occur in the homogeneous, but anisotropic case.

\section{Fulfillment of the Hadamard condition by the SLE}\label{SLE-Hadamard}

We will now show that the SLE are Hadamard states. As stated earlier in this paper, Hadamard states are completely characterized by the singularity structure of their two-point function, which means that the difference between the two-point functions corresponding to different Hadamard states must be a smooth function. Therefore we will compare the two-point function corresponding to the SLE to another one, corresponding to a given Hadamard state, and check that their difference is smooth. For this purpose, we will use the concept of adiabatic states, which are known to be, under certain conditions, Hadamard states. Moreover, this will give us an explicit ansatz for $T_{j}(t)$. In order to verify that the SLE satisfy the Hadamard condition, we will need a refinement of the notion of wavefront sets.

On the following, we will first introduce this refined notion of wavefront set and the definition of adiabatic states in terms of this notion. In the sequel, we will present the iteration procedure which provides the explicit ansatz for $T_{j}(t)$. After that we will use these as tools to show that the SLE constructed in the former section satisfy the Hadamard condition.

\subsection{Adiabatic States} \label{adiabaticstates}$ $

A distribution $u \in \D '(\mathbb{R}^{n})$ is said to be {\it microlocally} $H^{s}$ at $(x,k) \in \mathbb{R}^{n}\times(\mathbb{R}^{n}\diagdown\{0\})$ if there exists a conic neighborhood $V$ of $k$ and $\phi \in C_{0}^{\infty}(\mathbb{R}^{n})$, $\phi(x)\neq 0$, such that
\[\int_{V}\textrm{d}^{n}k\, \left(1+\lvert k \rvert^{2}\right)^{s}\lvert [\phi u]^{\wedge}(k) \rvert^{2}\leqslant \infty \; .\]

\begin{mydef}
 The Sobolev Wavefront set $WF^{s}$ of a distribution $u \in \D '(\mathbb{R}^{n})$ is the complement, in ${\mathcal T}^{*}\mathbb{R}^{n} \diagdown \{0\}$, of the set of all pairs $(x,k)$ at which $u$ is microlocally $H^{s}$.
\end{mydef}

Junker and Schrohe \cite{JunSchrohe02} showed that the Sobolev wavefront set of a distribution on any subset $X$ of $\mathbb{R}^{n}$ is a subset of ${\mathcal T}^{*}X \diagdown \{0\}$, and that, by choosing a suitable partition of unity, this definition can be extended for any paracompact smooth manifold $\M$. Besides, they proved the following Lemma:
\begin{Lemma}\label{Hadamard-Sobolev}
 For every Hadamard state $\omega_{H}$ we have
\begin{equation}
 WF^{s}(w^{(2)}_{\omega_{H}})=\left\{\begin{array}{cc}
                           \emptyset \textrm{\ ,\ } &s< -1/2 \\
			   C^{+} \textrm{\ ,\ } &s\geqslant -1/2
                          \end{array}
\right. \; ,
\end{equation}
\end{Lemma}
where $C^{+}$ is the set of points which composes the smooth wavefront set of a Hadamard state (see definition \eqref{Hadamard-wf} and equation \eqref{HadWF-C+}).

The adiabatic states are formulated iteratively (see below). For the $N-$th order of iteration, the adiabatic states $\omega_{N}$ are defined by the singularity structure of their two-point function:
\begin{mydef}
 A quasifree state $\omega_{N}$ on the $CCR$-algebra $\F$ is an {\it Adiabatic State} of order $N$ if, $\forall s<N+3/2$,
\begin{equation}
 WF^{s}(w^{(2)}_{\omega_{N}})=C^{+} \; .
 \label{Adiabatic-def}
\end{equation}
\end{mydef}

Comparing this with Lemma \eqref{Hadamard-Sobolev}, we have:
\begin{equation}
 WF^{s}(w^{(2)}_{\omega_{H}}-w^{(2)}_{\omega_{N}})=\emptyset \textrm{\quad ,\quad} \forall s<N+3/2 \; .
 \label{AdiabaticHadamard}
\end{equation}
Junker and Schrohe showed that the explicit construction given in \cite{LuRo90} satisfies the $WF^{s}$ condition. Furthermore, they defined adiabatic states on general globally hyperbolic spacetimes with compact Cauchy hypersurface. Hence the definition is also valid on the expanding spacetimes considered here. We will present this construction now.

The adiabatic ansatz determines the initial conditions of the solutions to the field equation \eqref{timeKG}. A solution to this equation, $S_{j}(t)$, assumes, as initial values,
\begin{equation}
 S_{j}(t_{0})=W_{j}(t_{0}) \textrm{\qquad ;\qquad} \dot{S}_{j}(t_{0})=\dot{W}_{j}(t_{0}) \; .
\end{equation}
At a generic instant of time,
\begin{equation}
 S_{j}(t)=\varsigma_{j}(t)W_{j}(t)+\xi_{j}(t)\overline{W}_{j}(t) \; .
 \label{Adansatz}
\end{equation}
$W_{j}$ is of the WKB form:
\begin{equation}
 W_{j}(t)=\frac{1}{\sqrt{2\Omega_{j}(t)c(t)^{3}}}\exp\left(i\int_{t_{0}}^{t}\textrm{d}t'\, \Omega_{j}(t')\right) \; .
 \label{WKB}
\end{equation}
$\Omega_{j}(t)$ is determined iteratively:
\begin{align}
 \Omega_{j}^{(0)} &=\omega_{j} \nonumber \\
 (\Omega_{j}^{(N+1)})^{2} &=\omega_{j}^{2}-\frac{3(\dot{c})^{2}}{4c^{2}}-\frac{3\ddot{c}}{2c}+\frac{3(\dot{\Omega}_{j}^{(N)})^{2}}{4(\Omega_{j}^{(N)})^{2}}-\frac{\ddot{\Omega}_{j}^{(N)}}{2\Omega_{j}^{(N)}} \; .
 \label{WKBiteration}
\end{align}

In the formulas appearing below, whenever a superscript $^{(N)}$ is present, it is meant that we are considering the $N$-th order of the adiabatic iteration. The authors of \cite{LuRo90} showed that, within a certain interval of time $\mathscr{I}$ and for large values of $\lambda_{j}$ (the ``frequency'' $\omega_{j}$ is defined in terms of the eigenvalues of the laplacian $\lambda_{j}$ in equation \eqref{frequency}), there exist constants $C_{\xi} \, ,\, C_{\varsigma}>0$ such that
\begin{align}
 \Omega_{j}^{(N)}(t)= &\mathcal{O}((1+\lambda_{j})^{1/2}) \; , \nonumber \\
\lvert  \xi_{j}^{(N)}(t) \rvert \leqslant C_{\xi}(1+\lambda_{j})^{-N-1/2} \; \; &, \; \; \lvert  1-\varsigma_{j}^{(N)}(t) \rvert \leqslant C_{\varsigma}(1+\lambda_{j})^{-N-1/2} \; ,
\label{WKBpar-j}
\end{align}
and
\begin{equation}
 \lvert W_{j}^{(N)}(t) \rvert =\mathcal{O}((1+\lambda_{j})^{-1/4}) \; \; \textrm{and} \; \; \lvert \dot{W}_{j}^{(N)}(t) \rvert =\mathcal{O}((1+\lambda_{j})^{1/4}) \; .
 \label{WKBsol-j}
\end{equation}

Now we will show that $WF^{s}(w^{(2)}_{\omega_{SLE}}-w^{(2)}_{\omega_{N}})=\emptyset$ and, by property \eqref{WFsum}, we will have $WF^{s}(w^{(2)}_{\omega_{SLE}}-w^{(2)}_{\omega_{H}})=\emptyset$.

\subsection{Fulfillment of conditions} \label{subsecfulfillhadamard}$ $

From \eqref{Adansatz}, \eqref{WKBpar-j} and \eqref{WKBsol-j},
\begin{equation}
 \partial_{t}^{k}S_{j}^{(N)}(t)=\mathcal{O}((1+\lambda_{j})^{k/2-1/4}) \; .
 \label{Ad-J}
\end{equation}

The two-point function corresponding to the SLE is given by \eqref{Bogstate}, where $c_{1j}$, $c_{2j}$ and $\beta_{j}$ are given by \eqref{c1j}, \eqref{c2j} and \eqref{Bogmin}, respectively. Since $c_{1j} > \lvert c_{2j} \rvert $,
\begin{equation}
 2\beta_{j}^{2} \approx \frac{1}{2}\frac{\lvert c_{2j} \rvert^{2}}{c_{1j}^{2}}+\frac{1}{4}\frac{\lvert c_{2j} \rvert^{4}}{c_{1j}^{4}}+\ldots 
\end{equation}

From \eqref{c1j} and \eqref{Ad-J}, it is immediate to see that
\begin{equation}
 c_{1j}=\mathcal{O}((1+\lambda_{j})^{1/2}) \; .
 \label{c1j-J}
\end{equation}
The analysis of the behavior of $\lvert c_{2j} \rvert $ is more involved. For this we need to estimate the scalar products of the WKB functions. The first one already appeared in equation \eqref{WKBsol-j}:
\begin{equation}
\left(W_{j}^{(N)},W_{j}^{(N)}\right)=\int_{I} dt\frac{1}{2c(t)\Omega_{j}^{(N)}(t)}=\mathcal{O}((1+\lambda_{j})^{-1/4}) \; .
\end{equation}
On the other hand, the scalar product
\begin{equation}
\left(\overline{W}_{j}^{(N)},W_{j}^{(N)}\right)=\int_{I}dt\frac{1}{2c(t)\Omega_{j}^{(N)}(t)}\exp{2i\int_{t_0}^t\Omega_{j}^{(N)}(t')dt'}
\end{equation}
is rapidly decaying in $\lambda_j$. This follows from the stationary phase approximation. It can be directly seen by exploiting the identity
\[\exp{2i\int_{t_0}^t\Omega_j^{(N)}(t')dt'}=\frac{1}{2i\Omega_j^{(N)}(t)}\frac{\partial}{\partial t}\exp{2i\int_{t_0}^t\Omega_j^{(N)}(t')dt'}\]
several times and subsequent partial integration. The estimates on $\Omega_j^{(N)}$ and its derivatives, together with the smoothness of $c(t)$, then imply the claim. Using these results, we have $\lvert c_{2j}^{(N)} \rvert =\mathcal{O}(\lambda_{j}^{-N})$. Therefore,
\begin{equation}
 \beta_{j}^{(N)}=\mathcal{O}(\lambda_{j}^{-N-1/2}) \; .
 \label{Bog-J}
\end{equation}

Now, we need similar estimates for the eigenfunctions and eigenvalues of the laplacian. The asymptotic behavior of the eigenvalues is directly given by Weyl's estimate \cite{Jost11}:
\begin{equation}
\lambda_{j}=\mathcal{O}(j^{2/m}) \; ,
\label{lambda-J}
\end{equation}
where $m$ is the dimension of the Riemannian manifold.

For the estimate on $\psi_{j}$, we start by defining the spectral function of the Laplace operator as the kernel of the projection operator on the subspace of all eigenfunctions of the Laplace operator whose corresponding eigenvalues are smaller than a certain value $\lambda$:
\begin{equation}
 e(\uline{x},\uline{y},\lambda) \coloneqq \sum_{\lambda_{j}\leqslant\lambda}\psi_{j}(\uline{x})\overline{\psi}_{j}(\uline{y}) \; .
 \label{spectralfunction}
\end{equation}
Elliptic regularity guarantees that this sum is bounded. The author of \cite{Hormander-IV} proved that, for any differential operator $Q_{\uline{x},\uline{y}}$ of order $\mu$, the following inequality is valid:
\begin{equation}
 \lvert Q_{\uline{x},\uline{y}}(e(\uline{x},\uline{y},\lambda)) \rvert \leqslant C_{Q}\lambda^{m+\mu} \; ,
\end{equation}
where $m$ is the dimension of $\Sigma$. Combining this result with the Weyl's estimate and restricting to $m=3$, we obtain
\begin{equation}
 \lvert \partial^{\lvert k\rvert}\psi_{j}(\uline{x}) \rvert^{2} \leqslant C_{3,k}\lambda_{j}^{3+2\lvert k \rvert} \; \therefore \; \lvert \partial^{\lvert k\rvert}\psi_{j} \rvert = \mathcal{O}(j^{1+2\lvert k\rvert /3}) \; .
 \label{eigen-lapl-J}
\end{equation}

Now, we proceed to the proof that the SLE are Hadamard states. As stated at the beginning of this section, adiabatic states $\omega_{N}$ in spacetimes with metric \eqref{metric-timedecomp-GH} and compact Cauchy hypersurface are Hadamard states. To show that the SLE are Hadamard, it suffices to show that
\[w^{(2)}_{\omega_{SLE}}-w^{(2)}_{\omega_{N}} \in H^{s}(\M \times \M) \; ,\]
for $s<N+3/2$. Moreover, since
\[C^{k}(\M \times \M) \subset H^{s}(\M \times \M) \textrm{\quad} \forall s<k-\frac{1}{2}\dim(\M \times \M) \; ,\]
all that is needed is to show that $\exists \, k>0$ such that
\begin{equation}
 w^{(2)}_{\omega_{SLE}}-w^{(2)}_{\omega_{N}} \in C^{k}(\M \times \M) \; .
 \label{SLE-Hadamard-cond}
\end{equation}

The difference between the two-point functions is given by
\begin{equation}
 (w^{(2)}_{\omega_{SLE}}-w^{(2)}_{\omega_{N}})(t,\uline{x};t',\uline{x'})=\sum_{j}\left(\overline{T}_{j}(t)T_{j}(t')-\overline{S}_{j}^{(N)}(t)S_{j}^{(N)}(t')\right)\psi_{j}(\uline{x})\overline{\psi}_{j}(\uline{x'}) \; .
 \label{SLE-N}
\end{equation}
We will verify the convergence of this sum by estimating the asymptotic behavior of its derivatives:
\begin{equation}
 \partial_{x,x'}^{\lvert k\rvert}(w^{(2)}_{\omega_{SLE}}-w^{(2)}_{\omega_{N}})(x;x')=\sum_{j}\partial_{x,x'}^{\lvert k\rvert}\left[\left(\overline{T}_{j}(t)T_{j}(t')-\overline{S}_{j}^{(N)}(t)S_{j}^{(N)}(t')\right)\psi_{j}(\uline{x})\overline{\psi}_{j}(\uline{x'})\right] \; .
 \label{partial-SLE-N}
\end{equation}

Since $T_{j}(t)$ is obtained from the adiabatic ansatz, it should be viewed as $T_{j}^{(N)}(t)$. Performing the Bogolubov transformation to write the two-point function of the SLE as \eqref{Bogstate}, we find
\begin{align}
 \overline{T}_{j}^{(N)}(t)T_{j}^{(N)}(t')-\overline{S}_{j}^{(N)}(t)S_{j}^{(N)}(t') &=(\beta_{j}^{(N)})^{2}\left[S_{j}^{(N)}(t)\overline{S}_{j}^{(N)}(t')+S_{j}^{(N)}(t')\overline{S}_{j}^{(N)}(t)\right] \nonumber \\
 &+2\beta_{j}^{(N)}\sqrt{1+(\beta_{j}^{(N)})^{2}}\textrm{Re}\left[e^{i\theta_{j}}S_{j}^{(N)}(t)S_{j}^{(N)}(t')\right] \; .
 \label{time-SLE-N}
\end{align}
From \eqref{Ad-J} and \eqref{Bog-J}, the last term on the rhs of \eqref{time-SLE-N} has the largest order in $j$. For that reason, this is the only term which we will take into account in the verification of the convergence of the sum.

Rewriting the estimates \eqref{Ad-J} and \eqref{Bog-J} in terms of $j$, we have
\begin{align}
 \partial_{t}^{k}S_{j}^{(N)}(t) &=\mathcal{O}(j^{k/3-1/6}) \label{S-J} \; , \\
 \beta_{j}^{(N)} &=\mathcal{O}(j^{-2N/3-1/3}) \; . \label{beta-J}
\end{align}

It is then easy to see that the derivative of largest order in \eqref{partial-SLE-N} is $\partial_{\uline{x},\uline{x'}}^{\lvert k\rvert}$:
\begin{equation}
 \partial_{\uline{x},\uline{x'}}^{\lvert k\rvert}\left[\left(\overline{T}_{j}(t)T_{j}(t')-\overline{S}_{j}^{(N)}(t)S_{j}^{(N)}(t')\right)\psi_{j}(\uline{x})\overline{\psi}_{j}(\uline{x'})\right]=\mathcal{O}\left(j^{\frac{4\lvert k\rvert}{3}-\frac{2N}{3}+\frac{4}{3}}\right) \; .
\end{equation}

The sum in \eqref{partial-SLE-N} will be absolutely convergent if
\begin{equation}
 \frac{4\lvert k\rvert}{3}-\frac{2N}{3}+\frac{4}{3}<-1 \; \therefore \; \lvert k\rvert<\frac{N}{2}-\frac{7}{4} \; .
\end{equation}
This means that
\begin{equation}
 w^{(2)}_{\omega_{SLE}}-w^{(2)}_{\omega_{N}} \in C^{\lfloor \frac{N}{2}-\frac{7}{4} \rfloor}(\M \times \M) \; ,
\end{equation}
where
\begin{equation}
 \lfloor x \rfloor \coloneqq \left\{\begin{array}{cc}
                           \max\{m \in \mathbb{Z}|m \leqslant x\} \textrm{\ ,\ } & x>0 \\
			   0 \textrm{\ ,\ } &x \leqslant 0
                          \end{array}
\right. \; .
\end{equation}

Finally,
\begin{equation}
 WF^{s}(w^{(2)}_{\omega_{SLE}}-w^{(2)}_{\omega_{N}})=\emptyset \textrm{\quad for \quad}s<\frac{N}{2}-\frac{23}{4} \; .
 \label{WF-SLE-N}
\end{equation}
Since $N+\frac{3}{2}>\frac{N}{2}-\frac{23}{4}$, the above equality means that $\forall s>-1/2\, ,\, \exists N\in \mathbb{Z_{+}}$ such that the adiabatic states are Hadamard states and, at the same time, satisfy \eqref{WF-SLE-N}\footnote{The equality \eqref{WF-SLE-N} is valid $\forall s\in \mathbb{R}$, but for $s<-1/2$ the Sobolev wavefront set of a Hadamard state is itself empty.} . Therefore,
\begin{equation}
 WF(w^{(2)}_{\omega_{SLE}}-w^{(2)}_{\omega_{H}})=\emptyset \; .
 \label{SLE-Hadamard-fulfilled}
\end{equation}

This proves that the States of Low Energy constructed on globally hyperbolic spacetimes with metric of the form \eqref{metric-timedecomp-GH} and compact Cauchy hypersurface are Hadamard states. We remark that this proof is also valid for the SLE constructed on homogeneous spacetimes above.

\section{Conclusions}

We have constructed states on expanding spacetimes without spatial symmetries which, at the same time, satisfy the Hadamard condition and minimize the expectation value of the smeared energy density. Our construction requires that the smearing is performed over a spatially extended region, a necessity which was not present in the original definition of SLE. However, we have generalized a procedure which was valid only for symmetric spacetimes.

Despite this definition of SLE being more general, we do not claim that this construction is the most general possible, because, generically, the metric on a globally hyperbolic spacetime is of the form \eqref{metric-GH}. Our construction relies on the existence of time-independent modes. A more general construction would require different techniques that do not rely on the ocurrence of such modes, such as an analysis based on pseudo-differential calculus, as the authors of \cite{GerardWrochna12} made in order to construct Hadamard states. Furthermore, the states here defined depend on the particular test function chosen for the smearing. Although the particle production process on RW spacetimes was shown not to be strongly dependent on it \cite{Degner09}, the complete role played by the test function on observational results is not yet well understood.

The existence of Hadamard states on a general globally hyperbolic spacetime is long known \cite{FullingNarcowichWald81}, although it was proven in a rather indirect way. A lot of work has been devoted to the construction of Hadamard states on cosmological spacetimes. Besides the adiabatic states, one has the so-called Bunch-Davies vacuum state on de Sitter spacetime \cite{Allen85}, whose formulation was recently generalized to asymptotically de Sitter spacetimes \cite{DappiaggiMorettiPinamonti09}. We remark that these states differ from the presently constructed ones since, in their case, whenever the spacetime possesses an everywhere timelike Killing vector field, the one parameter group which implements the action of this vector field on the GNS representation associated to that state has a positive self-adjoint generator. In the coordinate system used in the present work, this would amount to $\beta_{j}=0$. Another recent construction of Hadamard states was given in \cite{BrumFredenhagen13}, where the authors construct Hadamard states based on the spectral decomposition of the causal propagator on relatively compact spacetimes. These also differ from the present ones. For a timely review, including KMS states, almost KMS and almost equilibrium states on static and cosmological spacetimes, see \cite{BeniniDappiaggiHack13}.

\ack

Part of this work was based on the diploma thesis of Kolja Them \cite{Them10}, which was done under the supervision of Prof. Klaus Fredenhagen. Kolja Them would like to thank Prof. Fredenhagen, Thomas Hack and (former) members of the AQFT group at the University of Hamburg for several discussions. Marcos Brum would like to thank the AQFT group at the University of Hamburg for the hospitality during the period of preparation of this work. Marcos Brum is also indebted to Mr Zhirayr Avetisyan from the Max Planck Institute for Mathematics in the Sciences for enlightening discussions about Harmonic Analysis. Marcos Brum acknowledges financial support from the Brazilian agency CAPES under Grant nr 869211-4.

\section*{References}

\end{document}